\documentclass[preprint,showpacs,superscriptaddress,preprintnumbers]{revtex4}

\usepackage{graphicx}
\usepackage{dcolumn}
\usepackage{bm}
\usepackage{color} %

\begin{document}

\preprint{APS/123-QED}

\title{Emergent topological spin structures in a centrosymmetric cubic perovskite
}

\author{S. Ishiwata}
 \email{ishiwata@ap.t.u-tokyo.ac.jp}
 \affiliation{Department of Applied Physics and Quantum-Phase Electronics Center (QPEC), University of Tokyo, Hongo, Tokyo 113-8656, Japan}
 \affiliation{JST-PRESTO, Kawaguchi, Saitama 332-0012, Japan}

\author{T. Nakajima}%
\affiliation{RIKEN Center for Emergent Matter Science (CEMS), Wako, Saitama 351-0198, Japan}%

\author{J. -H. Kim}%
\affiliation{Max-Planck-Institut f$\ddot{u}$r Festk$\ddot{o}$rperforschung, D-70569 Stuttgart, Germany}%

\author{D. S. Inosov}%
\affiliation{Max-Planck-Institut f$\ddot{u}$r Festk$\ddot{o}$rperforschung, D-70569 Stuttgart, Germany}%
\affiliation{Institut f\"ur Festk\"orper- und Materialphysik, TU Dresden, 01069 Dresden, Germany}%

\author{N. Kanazawa}%
\affiliation{Department of Applied Physics and Quantum-Phase Electronics Center (QPEC), University of Tokyo, Hongo, Tokyo 113-8656, Japan}%

\author{J. S. White}%
\affiliation{Laboratory for Neutron Scattering and Imaging (LNS), Paul Scherrer Institut (PSI), CH-5232 Villigen, Switzerland}%

\author{J. L. Gavilano}%
\affiliation{Laboratory for Neutron Scattering and Imaging (LNS), Paul Scherrer Institut (PSI), CH-5232 Villigen, Switzerland}%

\author{R. Georgii}%
\affiliation{Physik Department E21, Technische Universit$\ddot{a}$t M$\ddot{u}$nchen, D-85748 Garching, Germany}
\affiliation{Heinz Maier-Leibnitz Zentrum (MLZ), Technische Universit$\ddot{a}$t M$\ddot{u}$nchen, D-85748 Garching, Germany}%

\author{K. Seemann}%
\affiliation{Heinz Maier-Leibnitz Zentrum (MLZ), Technische Universit$\ddot{a}$t M$\ddot{u}$nchen, D-85748 Garching, Germany}%

\author{G. Brandl}%
\affiliation{Heinz Maier-Leibnitz Zentrum (MLZ), Technische Universit$\ddot{a}$t M$\ddot{u}$nchen, D-85748 Garching, Germany}%

\author{P. Manuel}%
\affiliation{ISIS Facility, STFC Rutherford Appleton Laboratory, Chilton, Didcot, Oxfordshire, OX11 0QX, United Kingdom}%

\author{D. D. Khalyavin}%
\affiliation{ISIS Facility, STFC Rutherford Appleton Laboratory, Chilton, Didcot, Oxfordshire, OX11 0QX, United Kingdom}%

\author{S. Seki}%
\affiliation{RIKEN Center for Emergent Matter Science (CEMS), Wako, Saitama 351-0198, Japan}%

\author{Y. Tokunaga}%
\affiliation{Department of Advanced Materials Science, University of Tokyo, Kashiwa 277-8561, Japan}%

\author{M. Kinoshita}%
\affiliation{Department of Applied Physics and Quantum-Phase Electronics Center (QPEC), University of Tokyo, Hongo, Tokyo 113-8656, Japan}%

\author{Y. W. Long}%
\affiliation{RIKEN Center for Emergent Matter Science (CEMS), Wako, Saitama 351-0198, Japan}
\affiliation{Beijing National Laboratory for Condensed Matter Physics, Institute of Physics, Chinese Academy of Sciences, Beijing 100190,  China}%

\author{Y. Kaneko}%
\affiliation{RIKEN Center for Emergent Matter Science (CEMS), Wako, Saitama 351-0198, Japan}%

\author{Y. Taguchi}%
\affiliation{RIKEN Center for Emergent Matter Science (CEMS), Wako, Saitama 351-0198, Japan}%

\author{T. Arima}%
\affiliation{RIKEN Center for Emergent Matter Science (CEMS), Wako, Saitama 351-0198, Japan}
\affiliation{Department of Advanced Materials Science, University of Tokyo, Kashiwa 277-8561, Japan}%

\author{B. Keimer}%
\affiliation{Max-Planck-Institut f$\ddot{u}$r Festk$\ddot{o}$rperforschung, D-70569 Stuttgart, Germany}%

\author{Y. Tokura}%
\affiliation{Department of Applied Physics and Quantum-Phase Electronics Center (QPEC), University of Tokyo, Hongo, Tokyo 113-8656, Japan}%
\affiliation{RIKEN Center for Emergent Matter Science (CEMS), Wako, Saitama 351-0198, Japan}%

\date{\today}

\begin{abstract}
The skyrmion crystal (SkX) characterized by a multiple-$\bm{q}$ helical spin modulation has been reported as a unique topological state that competes with the single-$\bm{q}$ helimagnetic order in noncentrosymmetric materials. Here we report the discovery of a rich variety of multiple-$\bm{q}$ helimagnetic spin structures in the centrosymmetric cubic perovskite SrFeO$_3$. On the basis of neutron diffraction measurements, we have identified two types of robust multiple-$\bm{q}$ topological spin structures that appear in the absence of external magnetic fields: an anisotropic double-$\bm{q}$ spin spiral and an isotropic quadruple-$\bm{q}$ spiral hosting a three-dimensional lattice of hedgehog singularities. The present system not only diversifies the family of SkX host materials, but furthermore provides an experimental missing link between centrosymmetric lattices and topological helimagnetic order. It also offers perspectives for integration of SkXs into oxide electronic devices.
\end{abstract}

\maketitle
The discovery of novel magnetic spin structures has the potential to open new fields in condensed matter physics. This is exemplified by magnetic skyrmions with a vortex-like spin configuration, which has led to a multitude of possible applications of topological spin textures in spintronics \cite{Bogdanov, Robler, Muhlbauer, Nagaosa_Tokura, Kanazawa_Adv}.
Highly symmetric crystal lattices allow magnetically ordered states with different equivalent propagation vectors $\bm{q}$, and complex mesoscopic superstructures can emerge from superpositions of several such degenerate states. The "skyrmion crystal" (SkX), a multi-$\bm{q}$ superposition of magnetic skyrmions, has garnered particular recent attention because of its intriguing connection to topological spin and charge transport phenomena \cite{MnSi_PRL, MnGe_PRL, Jonietz, Mochizuki}. 

So far, SkXs have mostly been reported for non-centrosymmetric lattices, with details depending on the symmetry of the underlying crystal lattice, the magnetic anisotropy, and the relative strength of the competing interactions; i.e., the ferromagnetic exchange interaction and the Dzyaloshinskii-Moriya (DM) interaction \cite{Kanazawa_Adv}. Three types of two-dimensional SkX characterized by multiple coplanar $\bm{q}$ vectors have been reported: (i) a Bloch-type SkX formed by superposing three proper-screw spin modulations, which has been found in chiral helimagnets such as B20 compounds (MnSi, FeGe, etc.)\cite{Muhlbauer,Yu_Nature,Munzer_FeCoSi,Yu_NMat}, Cu$_2$OSeO$_3$\cite{Seki}, and Co-Zn-Mn alloys \cite{Tokunaga}, (ii) a N\'eel-type SkX formed by three cycloidal spin modulations found in polar helimagnets, like GaV$_4$(S,Se)$_8$ and VOSe$_2$O$_5$\cite{Istevan,Fujima,Bordacs,Kurumaji}, and (iii) an antiskyrmion crystal formed by three spiral spin modulations found in a Mn-Pt-Sn inverse Heusler compound with $D2d$ symmetry \cite{Nayak}.
Recently, a three-dimensional topological spin structure generated by triple-$\bm{q}$ vectors that are orthogonal to each other has been tentatively identified in the B20 compound MnGe with a relatively strong DM interaction \cite{MnGe_PRB}. This spin structure has hedgehog and antihedgehog singularities, where the associated emergent magnetic monopole and antimonopole manifest themself as a source of anomalous magnetotransport phenomena \cite{MnGe_PRL, MnGe_NCom}. However, as the experiments were performed on polycrystalline samples, a detailed characterization of these structures has not been possible.

In noncentrosymmetric helimagnets, the DM interaction originating from spin-orbit coupling apparently plays an important role in the formation of a SkX by selecting both the helicity and vorticity for each skyrmion \cite{MnGe_PRL, MnGe_PRB, MnGe_NCom}. On the other hand, the emergence of SkXs has been theoretically predicted also to occur in magnetically frustrated centrosymmetric helimagnets with high lattice symmetry \cite{Okubo, Mostovoy_NCom, Wang, Batista, Ozawa, Hayami}. In the absence of the DM interaction, these helimagnets have the potential to show rich topological spin textures due to fewer constraints on the spin helix. There exist a large number of centrosymmetric helimagnets, some of which show multiple-$\bm{q}$ spin modulations such as the rare-earth magnets \cite{Forgan, Jensen}. However, the presence of topologically nontrivial helimagnetic phases in centrosymmetric systems remains to be explored. 

 The simple cubic perovskite SrFeO$_{3}$ with crystal structure displayed in Fig. 1(a) is known to host a helimagnetic order below 130 K with metallic conductivity \cite{MacChesney, Takeda_screw}. The origin of the helimagnetic order in SrFeO$_{3}$ and related iron oxides has been discussed in terms of the competition between the nearest-neighbor and the further-neighbor interactions \cite{Kim_PRL} or the double-exchange mechanism considering the itinerant oxygen $p$ holes \cite{Mostovoy, Mostovoy2017}. 
Early neutron diffraction data were described in terms of a single-$\bm{q}$ proper-screw spiral with propagation vector along [111] or equivalent directions of the cubic lattice \cite{Takeda_screw}. Recently, however, SrFeO$_{3}$ was shown to display a rich variety of helimagnetic phases depending on temperature and external magnetic field as shown in Fig. 1(a) \cite{Ishiwata_SFO, Reehuis}.  Among them, Phases I and II are extraordinary in the sense that they exhibit a large unconventional Hall effect \cite{Hayashi, Ishiwata_SFO, Chakraverty}. The presence of sharp phase transitions with unusual transport signatures indicates well-ordered magnetic superstructures, rather than an incoherent superposition of domains of the single-$\bm{q}$ structure with different propagation vectors. In both Phase I and Phase II, the Hall resistivity as a function of $\bm{H}$ along [111] increases nonlinearly and reaches a maximum below the phase boundary to Phase IV or Phase V.
While this behavior implies the emergence of noncoplanar and/or topological  spin textures with scalar spin chirality \cite{MnSi_PRL, MnGe_PRL}, the magnetic structures within each phase have remained elusive. 
In this work, on the basis of comprehensive single crystal neutron diffraction studies, we reveal that the magnetic structures of the mysterious Phases I and II in the centrosymmetric cubic perovskite SrFeO$_{3}$ are indeed topological in nature, being identified as anisotropic double-$\bm{q}$ and isotropic quadruple-$\bm{q}$ helimagnetic structures, respectively. 

Figures 1(b) and 1(c) illustrate the spin structures reproduced by the superposition of the double-$\bm{q}$ and quadruple-$\bm{q}$ magnetic modulations, which we propose in this study as models for Phases I and II, respectively. 
Note that we tentatively adopt the same helicity and phase for each spin modulation. Owing to the cubic symmetry of the crystal, there are four $\bm{q}$-vectors of $(q,q,q)$, $(-q,q,q)$, $(q,-q,q)$ and $(q,q,-q)$, where $q\sim$ 0.13; in this paper, we refer to them as $\bm{q}_1$, $\bm{q}_2$, $\bm{q}_3$ and $\bm{q}_4$, respectively, as described in Fig. 1(a). 
As explained later in detail, Phase I can be described as a double-$\bm{q}$ structure encompassing proper-screw and cycloidal modulations with slightly different $\bm{q}$-vectors, so that the overall symmetry of the superstructure is reduced (see supplementary information for details). Here we define the topological number as the integral of the solid angle made by the three adjacent spins around the singular point as described in Ref. \cite{Nagaosa_Tokura}. Following this definition, the anisotropic double-$\bm{q}$ spin spiral in Phase I turns out to be topologically nontrivial as shown in Fig. 1(b), where noncoplanar vortex-like spin configurations can be found. As for Phase II, 4 equivalent proper-screw-type spin modulations yield a face-centered cubic lattice of topological singularities, at which the hedgehog/anti-hedgehog spin texture acts as the source/sink of the emergent magnetic fields (Fig. 1(c)), i.e., an emergent magnetic monopole/antimonopole. The application of external magnetic fields changes the relative positions of the magnetic monopoles and antimonopoles, producing the effective magnetic flux necessary for the topological Hall effect\cite{MnGe_NCom}. 

First, let us identify the multiple-$\bm{q}$ state in Phases I and II on the basis of results of high-resolution neutron diffraction measurements for SrFeO$_3$ at the WISH (Wide angle In a Single Histogram) diffractometer. Figure 2(a) shows integrated intensities of incommensurate magnetic reflections near the $(1,\bar{1},0)$ reciprocal lattice point, which were measured on heating in zero field after field cooling (FC) with an external field $\bm{H}$ of 7 T along the [111] axis. The data labeled $\bm{q}_i$ show the intensity corresponding to the magnetic modulation vector $\bm{q}_i$. In the simple single-$\bm{q}$ proper-screw spin structure, there exist four kinds of domains in each of which only one of the $\bm{q}$-vectors is selected (the helicity is not considered here). If this were the case for Phase I, one of the domains with the propagation vector $\bm{q}_{1}$ parallel to $\bm{H}$ would be selected through the FC process due to the difference in the Zeeman energy, which would give rise to the nonvanishing scattering intensity only for $\bm{q}_{1}$. As seen in Fig. 2(a), however, nonzero scattering intensities are found not only for $\bm{q}_{1}$ but for $\bm{q}_{2-4}$. Thus, the possibility of the single-$\bm{q}$ proper-screw spin structure can be ruled out for Phase I and likewise for Phase II. It should be noted for Phase I that the scattering intensity for $\bm{q}_1$ is much larger than for $\bm{q}_{2-4}$. 
This is consistent with the anisotropic double-$\bm{q}$ spin structure, in which one of the $\bm{q}$-vectors with proper-screw-type spin spiral tends to be aligned along the external field, thus yielding a magnetic modulation with large amplitude. %
We found that all the scattering intensities are comparable to each other in Phase II, suggesting that the field-oriented anisotropic magnetic structure in Phase I disappears upon the first-order phase transition to Phase II. %

Having confirmed the multiple-$\bm{q}$ state in Phases I and II, we measured the $H$ dependence of the magnetic scattering intensity at the selected temperatures after zero-field cooling (ZFC) as shown in Fig. 2(b). At 50 K and zero field in Phase I, the scattering intensities for $\bm{q}_{1-4}$ are distributed in a certain range, reflecting the multidomain state of the anisotropic multiple-$\bm{q}$ spin structure. As $H$ is increased beyond 5 T, the intensity only for $\bm{q}_{1}$ parallel to $\bm{H}$ becomes larger, and the others become smaller. The significant $H$-induced change with a large hysteresis at low $H$ is ascribable to the domain reorientation, as also suggested by the previous report on the magnetoresistance anomaly measured after ZFC (see the shaded area in Figs. 1(a) and 2(b)) \cite{Ishiwata_SFO}. For Phase II, on the other hand, the scattering intensities for $\bm{q}_{1-4}$ are comparable to each other and $H$-dependent anomalies and hysteresis are absent (see Fig. 2(c)), being consistent with the presumed isotropic quadruple-$\bm{q}$ helimagnetic structure. Upon an increase in $H$ to 12 T that induces the II-IV phase transition, the intensity distribution tends to become the one expected for the single-$\bm{q}$ state. However, since the maximum $H$ of 12 T is located near the phase boundary and the scattering intensities for $\bm{q}_{2-4}$ remain nonzero, further experiments with larger $H$ are indispensable to characterize the spin structure of Phase IV.

To further characterize the multiple-$\bm{q}$ spin structure and the domain state in Phase I, we measured small-angle neutron scattering (SANS) for $\bm{q}_{1}$ and $\bm{q}_{2}$ after FC and ZFC. The SANS experiments were performed for SrFe$_{0.99}$Co$_{0.01}$O$_{3}$ having essentially the same phase diagram as SrFeO$_3$ (see Fig. 4(a)) \cite{Long}.
Figures 3(d) and 3(g) respectively show the magnetic reflections around $\bm{q}_{1}$ and $\bm{q}_{2}$ measured at 3 K in Phase I after FC. The scattering profile for $\bm{q}_{1}$ revealed that the magnetic modulation wave vector is no longer described by the simple $(q,q,q)$, but split into three peaks indexed as $(q,q,q')$ (= $\bm{q}_{1}'$), $(q,q',q)$ and $(q',q,q)$, where $|q'| > |q|$. On the other hand, a single peak is found for $\bm{q}_{2}$. However, this reflection is also slightly shifted from the [$\bar{1}\bar{1}$1] direction, and assigned as $(-q,-q,q')$ (= $\bm{q}_{2}'$). The slightly different $\bm{q}$-vector may reflect the influence of small, anisotropic terms in the spin Hamiltonian (see the discussion below). The observation of the triplet peaks around $\bm{q}_{1}$ after FC with $\bm{H}$ parallel to [111] is a signature of the three $\bm{q}$-dependent domains (with each domain containing four kinds of helicity) of the anisotropic double-$\bm{q}$-helimagnetic structure with $\bm{q}_{1}'$ and $\bm{q}_{2}'$. The azimuthal directions of these propagation vectors deviate from [111] and [$\bar{1}\bar{1}$1], respectively, so that Phase I can accommodate the angular mismatch between $\bm{q}_{1}'$ and $\bm{q}_{2}'$. In fact, we confirmed that the intensities for the magnetic scattering along the three $<$111$>$ equivalents, [1$\bar{1}\bar{1}$], [$\bar{1}1\bar{1}$], and [$\bar{1}\bar{1}$1], are significantly different even after the FC process, reflecting the imbalance of the three domains with the different $\bm{q}_{2}'$ directions (see Fig. S2). 
As the temperature increases through the I-II phase transition at 90 K, the directions of all propagation vectors become parallel to the $<$111$>$ equivalents, consistent with the proposal for the isotropic quadruple-$\bm{q}$ helmagnetic structure in Phase II. (see Figs. 3(d-i)). The schematic representations of the observed scattering peaks for Phases I and II are displayed in Figs. 3(a-c). The numbers of the domain types of Phase I in the ZFC and FC states are 48 and 12, respectively, the latter of which is obtained by considering the three-fold symmetry around [111], and the helicity degree of freedom associated with the two kinds of spin spirals. At least at low temperature, magnetic fields of order $\sim$10 T therefore mostly select different equivalent domains without modifying the magnetic structure substantially.

Next, we performed polarized SANS experiments to learn microscopically how the spins are twisting along [$\bar{1}\bar{1}$1], which is essentially equivalent to [$\bar{1}1\bar{1}$] and [1$\bar{1}\bar{1}$] with respect to the angle relative to $\bm{H}$. Here, we assume three possible arrangements of spin spirals: i) vertical-cycloid type with the spin-spiral plane parallel to $\bm{q}_{2}'$ and $\bm{H}$ (Fig. 4(d)); ii) proper-screw type with the spin-spiral plane normal to $\bm{q}_{2}$ (Fig. 4(e)); and iii) horizontal-cycloid type with the spin-spiral plane parallel to $\bm{q}_{2}$ and normal to $\bm{H}$ (Fig. 4(f)).  When considering that only the component of the magnetic moments perpendicular to the scattering vector causes neutron scattering, the effective spin components for each spin spiral can be described as shown at the right end of Figs. 4(d-f). Nevertheless, by using spin polarized SANS, we can distinguish between the three types of spin spiral. Since the non-spin-flip (NSF) and the spin-flip (SF) geometries detect only the spin components parallel and normal to $\bm{H}$, respectively, the intensity ratios of NSF and SF scattering at $\bm{q}_{2}'$ and $\bm{q}_{2}$ are expected to display the dependence represented by the relative lengths of the blue and red arrows in Figs. 4(d-f). The measurements were performed after FC in Phase I, so that 6 kinds of domains with $\bm{q}_{1}'$ nearly parallel to $\bm{H}$ are selected. The inset of Fig. 4(a) shows the $\omega$-scan profiles of the magnetic scattering for $\bm{q}_{2}'$ at 3 K and 0.3 T, which were normalized by the flipping ratios. As shown in Figs. 4(b) and 4(c), the normalized scattering intensity for the NSF geometry is much larger than that for the SF geometry in Phase I, whereas the scattering intensities for both geometries are nearly the same in Phase II. This result indicates that Phase I and Phase II encompass vertical-cycloid and proper-screw states, respectively, propagating along [$\bar{1}\bar{1}$1]. In Phase V at 7 T, the scattering intensity for the SF geometry is larger than that for the NSF geometry, implying that the spin arrangement propagating along [$\bar{1}\bar{1}$1] is in the horizontal-cycloid type. This result indicates that Phase V has multiple-$\bm{q}$ spin spirals as well, calling for further experiments to identify the spin structure. 

To summarize, we have identified two kinds of topological spin structures in SrFeO$_3$, that appear robustly even without external magnetic fields. The anisotropic double-$\bm{q}$ spin structure in Phase I manifests itself as a unique topological order reflecting the versatility of the centrosymmetric lattice that permits various types of spin spiral. However, this spin texture cannot explain the observed topological Hall resistivity, because the expected direction of the emergent magnetic flux is perpendicular to the external magnetic field. Future research will have to assess whether the discrepancy between the expected and observed directions of the emergent magnetic flux in Phase I arises from an additional internal spin modulation that is not resolved in the current experiment. On the other hand, Phase II can be described in terms of an isotropic quadruple-$\bm{q}$ spin spiral, that presumably yields emergent magnetic monopoles as reported for the noncentrosymmetric helimagnet MnGe \cite{MnGe_PRB}. 

We now turn to the mechanisms underlying the observed cascade of magnetic phase transitions. The fact that the diffraction patterns at the lowest temperature in ZFC and FC states are closely similar implies that Phase I is not stabilized by the Zeeman interaction, but rather by anisotropic terms in the zero-field spin Hamiltonian (such as magneto-crystalline anisotropies generated by the spin-orbit coupling) that go beyond the primary isotropic double-exchange and/or superexchange interactions. With increasing temperature, the corresponding free-energy gain could be offset by the higher entropy of the more symmetric quadruple-$\bm{q}$ structure, leading to the observed phase transition between Phases I and II. Recently, Monte Carlo simulations based on a Kondo lattice model with a biquadratic interaction defined in momentum space have indicated various multiple-$\bm{q}$ phases on a centrosymmetric lattice with rotational symmetry\cite{Hayami}. Although this model does not apply directly to our system, this theoretical work and our experimental results provide a timely showcase of the rich variety of multiple-$\bm{q}$ helimagnetic phases with topological singularities that can be expected to emerge ubiquitously in frustrated itinerant magnets with high lattice symmetry, even without the DM interaction. Moreover, perovskite-type oxides are a broad class of materials that already find many applications especially in the form of heterostructures enabling the interplay between topological magnetism and other collective quantum phenomena. The discovery of topological spin order in SrFeO$_3$ is therefore a milestone for integrating potential topological magnetic states into existing device architectures.  

\clearpage
\noindent 
\subsection*{APPENDIX: MATERIALS AND METHODS}

Single crystals of SrFeO$_3$ and SrFe$_{0.99}$Co$_{0.01}$O$_3$ were obtained by a high-pressure oxygen annealing for the large single crystals of the oxygen-deficient perovskite with brownmillerite-type structure as described in Refs. \cite{Ishiwata_SFO} and \cite{Long_JPCM}. The orientation of the single crystal with dimensions of about 3 $\times$ 3 $\times$ 2 mm$^3$ was checked by x-ray Laue diffractions.

The temperature and magnetic field variations of the neutron diffraction intensities shown in Fig. 2 were measured for SrFeO$_3$ with a cold neutron time-of-flight diffractometer, WISH (Wide angle In a Single Histogram) \cite{Chapon}, at the ISIS Facility in UK. The single crystal SrFeO$_3$ was loaded into a vertical-field cryomagnet, whose maximum field was 13.5 T. The cubic [111] axis was set to be parallel to the magnetic field. The cryomagnet has a large vertical opening angle from -5$^{\circ}$ to 10$^{\circ}$, which enabled us to measure the out-of-plane incommensurate magnetic reflections near the (1, $\bar{1}$, 0) reciprocal lattice point under magnetic field along the [111] direction. 
Small angle neutron scattering (SANS) measurements were carried out for SrFe$_{0.99}$Co$_{0.01}$O$_3$ with a vertical-field cryomagnet (7 T in maximum) at the MIRA beamline in FRM II and a horizontal-field cryomagnet (6.8 T in maximum) at the SINQ beamline in Paul Scherrer Institute \cite{Georgii1,Georgii2}. We employed an experimental setup of the horizontal configuration (see Fig. S1(a)), in which the sample can be rotated by 360$^{\circ}$ around the [$\bar{1}\bar{1}$2] axis ($\omega$ axis), and vertical configuration (see Fig. S1(b)), in which the sample can be rotated by 360° around the [111] axis ($\omega$ axis).  In both experimental configurations, magnetic field was applied parallel to the [111] axis and perpendicular to the incident neutron beam. Experimental data taken at MIRA were collected both in polarized and unpolarized modes at a wavelength $\lambda$ = 4.5 \AA, and those at SINQ were collected in unpolarized mode at $\lambda$ = 4.7 \AA. The polarization of neutron spins was generated by the magnetic field gradient near the vertical superconducting magnet. The flipping ratios for 0.3 T and 7 T are 4.9 and 3.0, respectively.

\clearpage
\noindent 
\begin{quote}

\end{quote}

\begin{figure} 
\begin{center}
\includegraphics[keepaspectratio,width=12 cm]{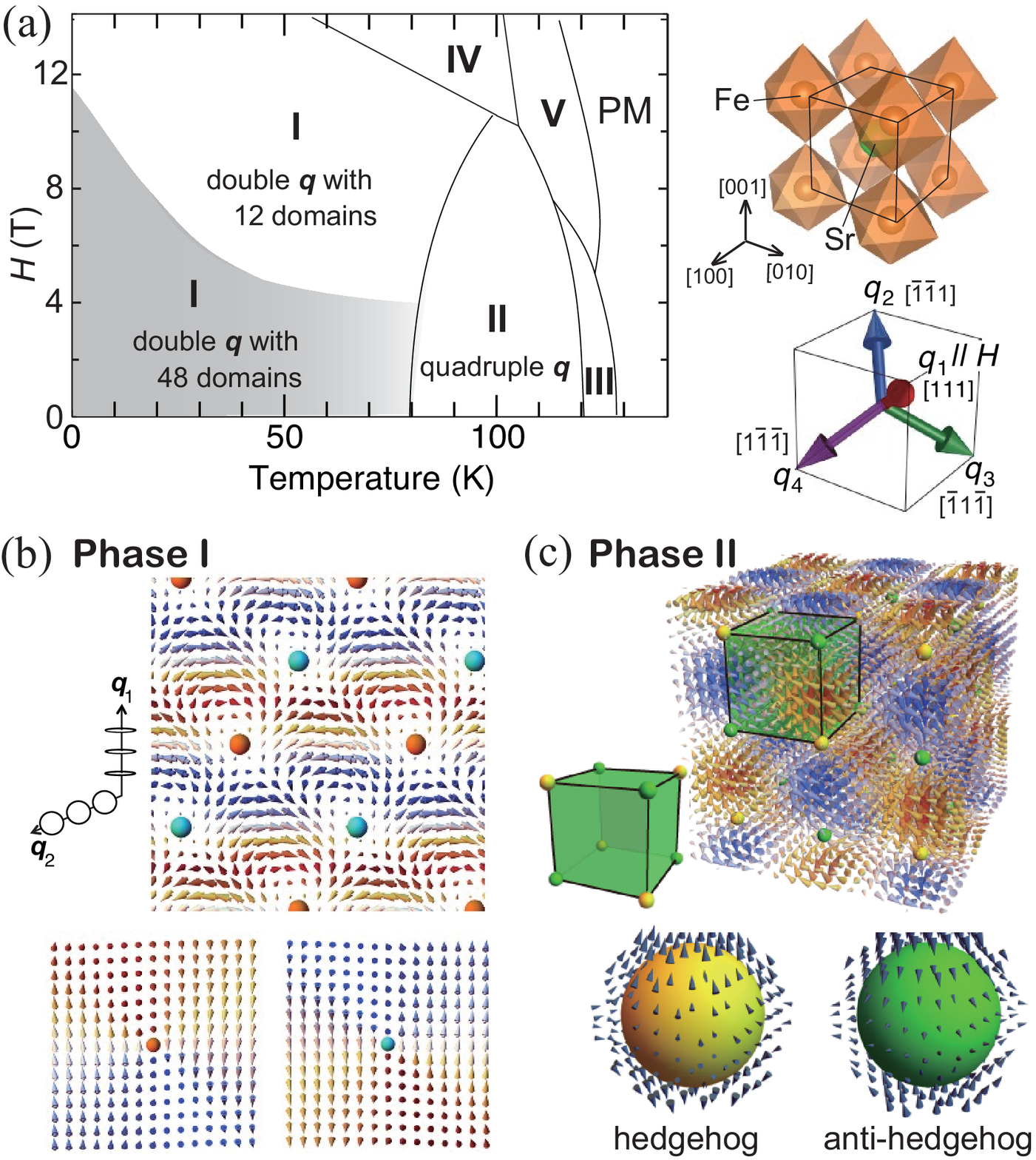}
\caption{\label{fig1} (a) Magnetic phase diagram for the applied field direction along [111]. The shaded and the white regions in Phase I correspond to states with 48 and 12 domains, respectively. Schematic crystal structure and quadruple-$\bm{q}$ vectors viewed along [111] are shown on the righthand side. (b) Double-$\bm{q}$ spin structure in Phase I and (c) quadruple-$\bm{q}$ spin structure in Phase II (the color of each spin corresponds to the spin component along the direction perpendicular to both $\bm{q}_{1}$ and $\bm{q}_{2}$ for panel (b) and that along [111] for panel (c), respectively). The magnified views around the singular points are shown at the bottom. Note that we adopt $\bm{q}_{1}$ and $\bm{q}_{2}$ instead of $\bm{q}'_{1}$ and $\bm{q}'_{2}$ for Phase I.}
\end{center}
\end{figure}

\begin{figure} 
\begin{center}
\includegraphics[keepaspectratio,width=10 cm]{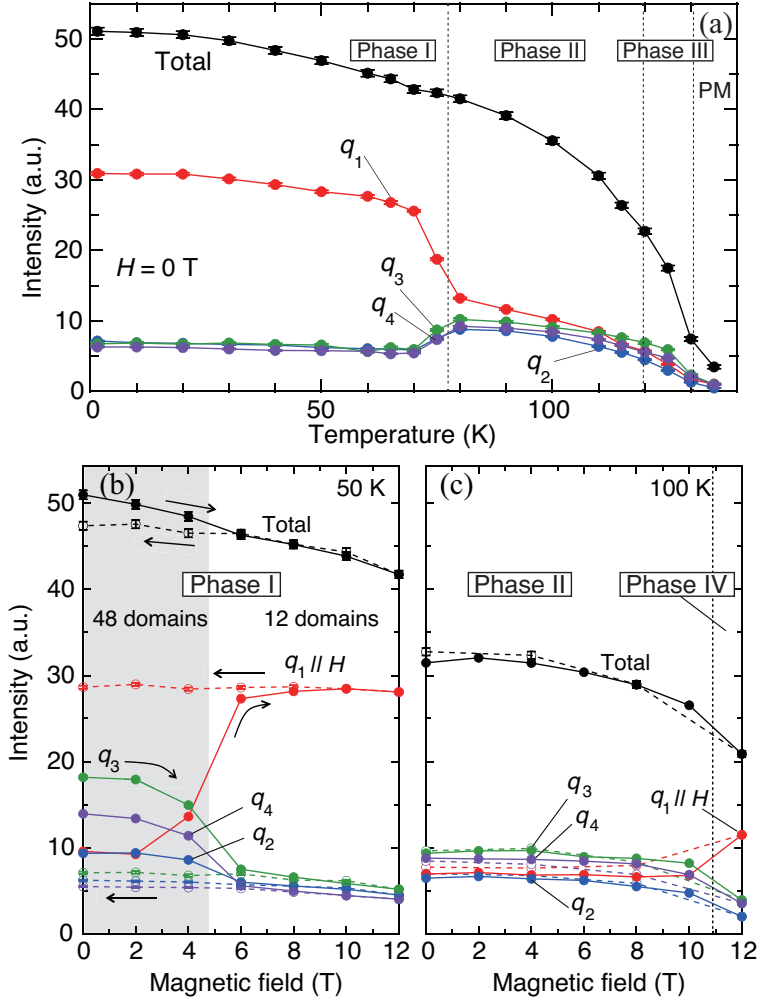}
\caption{\label{fig2} (a) Temperature dependence of integrated intensities of the magnetic scattering along $\bm{q}_i$ ($i=1-4$) measured at zero field after field cooling. The total intensity for $\bm{q}_{1-4}$ is also shown. Magnetic field dependence of the magnetic scattering intensities for (b) Phase I at 50 K and (c) Phase II at 100 K. The data with solid lines and broken lines were measured on increasing and decreasing the field along [111], respectively. All the data were measured after zero-field cooling from room temperature.} 
\end{center}
\end{figure}

\begin{figure} []
\begin{center}
\includegraphics[keepaspectratio,width=14 cm]{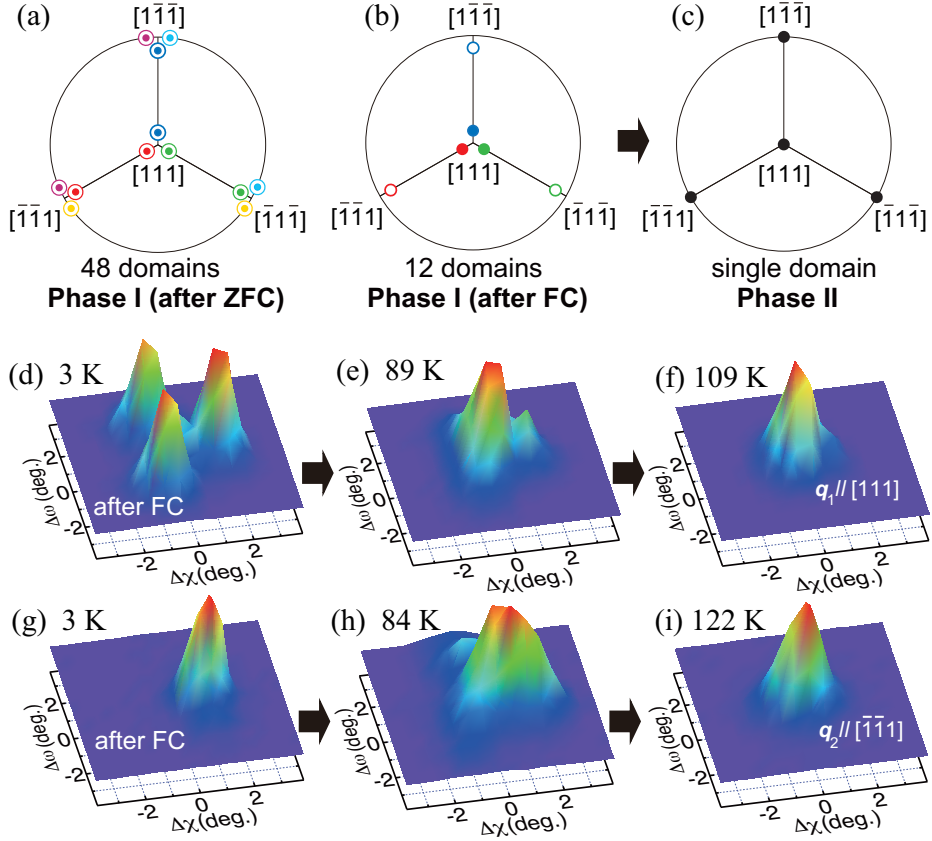}
\caption{\label{fig3} (a)-(c) Schematic representations of neutron scattering peaks for the spin spiral propagating along the four $\{$111$\}$ equivalents, which are supposed to appear at zero field after zero-field cooling in Phase I (a) and field cooling in Phase I (b) and Phase II (c). The filled and open circles represent the magnetic reflections for the proper-screw and the vertical-cycloid-type spin propagation, respectively. Open circles with cores in panel (a) correspond to overlapping magnetic reflections of the proper-screw and the vertical-cycloid types in a multidomain state. (b)-(f) Magnetic scattering profiles around $\bm{q}_{1}$ parallel to [111]. The peaks in panels $\bf{d}$ and $\bf{f}$ correspond to the triplet filled circles and a single open circle, respectively, in the panel $\bf{b}$. (g)-(i) Magnetic scattering profiles around $\bm{q}_{2}$ parallel to [$\bar{1}\bar{1}$1]. The intensity is normalized by the largest value. The data were collected on heating in zero field after field cooling under a magnetic field of 6.8 T for SrFe$_{0.99}$Co$_{0.01}$O$_{3}$. For the experimental setup, see Fig. S1.} 
\end{center}
\end{figure}

\begin{figure} []
\begin{center}
\includegraphics[keepaspectratio,width=14 cm]{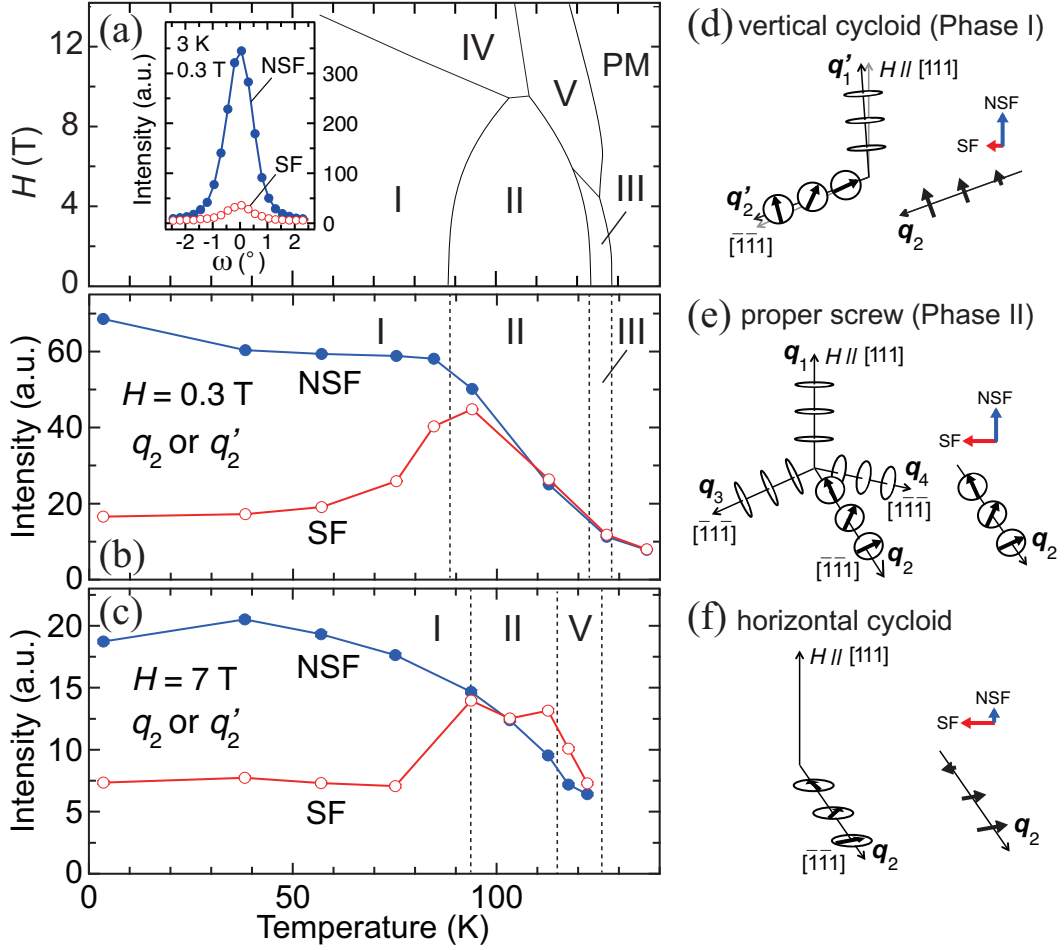}
\caption{\label{fig4} (a) The magnetic phase diagram of SrFe$_{0.99}$Co$_{0.01}$O$_{3}$ reproduced from \cite{Long}. The neutron scattering profiles at 3 K and 0.3 T for spin flip (SF) and non spin flip (NSF) geometries are shown as an inset. All the data were collected after field cooling in a magnetic field of 7 T. Temperature dependence of SF and NSF scattering intensities measured on heating at (b) 0.3 T and (c) 7 T. In each case, the applied magnetic field and the incident neutron spin are parallel to $[111]$. Schematic illustrations of the spiral spin propagation for $\bm{q}_{2}'$ or $\bm{q}_{2}$ with (d) vertical-cycloid-type (Phase I), (e) proper-screw-type (Phase II), and (f) horizontal-cycloid-type configurations, the right-hand side of which shows the effective spin components for spin-polarized SANS.}
\end{center}
\end{figure}



\bibliography{scibib}

\bibliographystyle{Science}

\clearpage

\section*{Acknowledgments}
The authors thank M. Mostovoy, S. Hayami, Y. Motome, N. Nagaosa, and M. Takano for useful comments, and thank N. Egetenmeyer for her kind experimental support. This work is partly supported by Grant-in-Aid for Scientific Research, Japan Society for the Promotion of Science, Japan (Kakenhi No. 17H01195 and No. 16K17736), JST PRESTO Hyper-nano-space design toward Innovative Functionality (Grant No. JPMJPR1412), and Asahi Glass Foundation. D. S. I. acknowledges support from the German Research Foundation (DFG) through the Collaborative Research Center SFB 1143 in Dresden (project C03). J. S. W. acknowledges support from Swiss National Science Foundation (SNSF) via the Sinergia network 'NanoSkyrmionics' (grant CRSII5-171003), and the SNSF project grant No. 153451. B.K. acknowledges support from the DFG through the Collaborative Research Center TRR80.



\end{document}